\documentclass[sigconf]{acmart}

\copyrightyear{2023}
\acmYear{2023}
\setcopyright{rightsretained}
\acmConference[FAccT '23]{2023 ACM Conference on Fairness, Accountability, and Transparency}{June 12--15, 2023}{Chicago, IL, USA}
\acmBooktitle{2023 ACM Conference on Fairness, Accountability, and Transparency (FAccT '23), June 12--15, 2023, Chicago, IL, USA}
\acmDOI{10.1145/3593013.3594112}
\acmISBN{979-8-4007-0192-4/23/06}

\usepackage{balance}
\usepackage{wrapfig}
\usepackage{subcaption}
\usepackage{xspace}
\usepackage{natbib}
\usepackage{hyperref}
\usepackage{algorithm}
\usepackage{enumitem}

\graphicspath{ {./images/} }

\DeclareMathOperator*{\argmax}{argmax}

\newcommand{\DiversitySet}{\ensuremath{\mathcal{D}}\xspace}
\newcommand{\metric}[2]{\ensuremath{\text{Div}@{#1}(#2)}\xspace}
\newcommand{\RankingSystem}{\ensuremath{R}\xspace}
\newcommand{\QuerySet}{\ensuremath{\mathcal{Q}\xspace}}
\newcommand{\kmax}{K_{\max}}

\newcommand{\DPPBatchSize}{\ensuremath{B}\xspace}
\newcommand{\DPPDepth}{\ensuremath{N'}\xspace}
\newcommand{\DPPWindowSize}{\ensuremath{w}\xspace}

\newcommand{\StrongOR}{\textsc{Strong-OR}\xspace}
\newcommand{\OverfetchAndRerank}{\textsc{Overfetch-and-Rerank}\xspace}

\newcommand{\DPPLong}{\textsc{Determinantal Point Process}\xspace}
\newcommand{\DPP}{\textsc{DPP}\xspace}
\newcommand{\RRLong}{\textsc{Round-Robin}\xspace}
\newcommand{\RR}{\textsc{RR}\xspace}
\newcommand{\BucketizedANN}{\textsc{Bucketized-ANN} Retrieval\xspace}
\newcommand{\MMRLong}{\textsc{Maximal Marginal Relevance}\xspace}
\newcommand{\MMR}{\textsc{MMR}\xspace}

\newcommand{\PDP}{Related Products\xspace}
\newcommand{\Search}{Search\xspace}
\newcommand{\Homefeed}{Homefeed\xspace}
\newcommand{\NUX}{New User Homefeed\xspace}

\begin{document}

\title[Practical End-to-End Diversification in Search and Recommender Systems]{Representation Online Matters: Practical End-to-End Diversification in Search and Recommender Systems}

\author{Pedro Silva}
\authornote{These authors contributed equally to this work.}
\email{psilva@pinterest.com}
\affiliation{%
  \institution{Pinterest, Inc.}
  \city{San Francisco}
  \state{California}
  \country{USA}
}

\author{Bhawna Juneja}
\authornotemark[1]
\email{bjuneja@pinterest.com}
\affiliation{%
  \institution{Pinterest, Inc.}
  \city{San Francisco}
  \state{California}
  \country{USA}
}

\author{Shloka Desai}
\authornotemark[1]
\email{sdesai@pinterest.com}
\affiliation{%
  \institution{Pinterest, Inc.}
  \city{San Francisco}
  \state{California}
  \country{USA}
}

\author{Ashudeep Singh}
\email{ashudeepsingh@pinterest.com} 
\affiliation{%
  \institution{Pinterest, Inc.}
  \city{San Francisco}
  \state{California}
  \country{USA}
}

\author{Nadia Fawaz}
\email{nadia.fawaz@acm.org}
\affiliation{%
  \institution{Pinterest, Inc.}
  \city{San Francisco}
  \state{California}
  \country{USA}
}

\begin{abstract}
As the use of online platforms continues to grow across all demographics, users often express a desire to feel represented in the content. 
To improve representation in search results and recommendations, we introduce end-to-end diversification, ensuring that diverse content flows throughout the various stages of these systems, from retrieval to ranking. We develop, experiment, and deploy scalable diversification mechanisms in multiple production surfaces on the \textit{Pinterest} platform, including \Search, \PDP, and \NUX, to improve the representation of different skin tones in beauty and fashion content. 
Diversification in production systems includes three components: identifying requests that will trigger diversification, ensuring diverse content is retrieved from the large content corpus during the retrieval stage, and finally, balancing the diversity-utility trade-off in a self-adjusting manner in the ranking stage. Our approaches, which evolved from using Strong-OR logical operator to bucketized retrieval at the retrieval stage and from greedy re-rankers to multi-objective optimization using determinantal point processes for the ranking stage, balances diversity and utility while enabling fast iterations and scalable expansion to diversification over multiple dimensions. Our experiments indicate that these approaches significantly improve diversity metrics, with a neutral to a positive impact on utility metrics and improved user satisfaction, both qualitatively and quantitatively, in production.
\end{abstract}


\maketitle

\section{Introduction}

Over the past decade, the use of online platforms has grown among all demographics and many communities have expressed the need to feel represented in content surfaced online \cite{pew2021}.  
While representation has gradually improved in some media \cite{nielsen2023}, it remains lacking on social media platforms and in search results and recommendations \cite{pew2018,kay2015unequal}. As technology becomes increasingly integrated into the daily lives of billions of people globally, it is crucial for online platforms to reflect the diverse communities they serve.  
Search and recommendation systems play a significant role in users' online experiences in various applications, from content discovery to entertainment and from e-commerce to media streaming. By paying close attention to the diversity reflected in their content, these systems can break away from historical patterns of bias and move towards a more inclusive and equitable online experience.

\begin{figure*}[t]
    \centering
    \includegraphics[width=\linewidth]{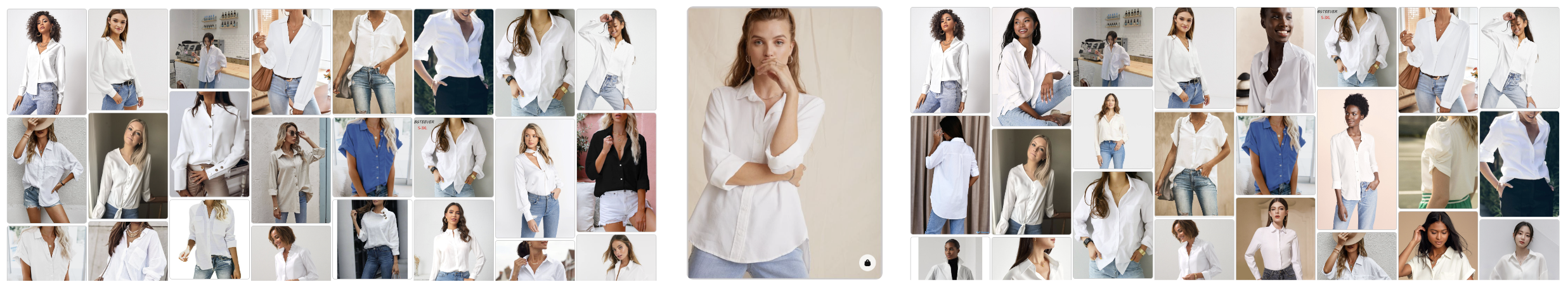}
    \caption{Side-by-side \PDP recommendations for the query Pin "Shirt Tail Button Down" shown in the center. Left: previous experience without diversification. Right: current diversified experience.}
    \Description{This image shows a side-by-side for Related Products recommendations for the query Pin "Shirt Tail Button Down", which is a woman wearing a white shirt and blue jeans. On the left we show the previous experience without diversification, which has a lot of pins with a white shirt but also some that have a blue shirt or a black shirt and a narrower range of skin tones. On the right is the diversified experience, in which all product pins have a white shirt and a wide variety of skintone ranges are represented.}
    \label{fig:pdp_diversified_small}
\end{figure*}

Improving representation online can facilitate content discovery for a more diverse user base by reflecting their inclusion on the platform. This, in turn, demonstrates the platform's ability to meet their needs and preferences.
In addition to improved user experience and satisfaction, this can have a positive business impact through increased engagement, retention, and trust in the platform. Gaining a deeper understanding of the diversity of user experiences and perspectives can also lead to a more diverse content corpus, which can significantly drive innovation and creativity. 

In this paper, we aim to address the challenge of diversification in large-scale search and recommender systems. Our focus is on diversification mechanisms for visual discovery on \textit{Pinterest}. Pinterest is the visual inspiration platform people all around the world use to discover the world’s most inspiring ideas, plan their best lives, and shop to make their plans a reality. Over 460 million users \cite{pinterestEarnings} use Pinterest monthly to discover ideas and products from a corpus of over 11 billion visual bookmarks called Pins. Pins can be images, videos, or products saved from the web or created by Pinners, creators, publishers, and businesses on the platform. People can search for Pins, save the ones they like and click on a Pin to visit a website and learn more. We focus on diversifying recommendations on three surfaces on Pinterest: \Search, \PDP recommendations, and \NUX.  Specifically, we develop, experiment, and deploy scalable diversification mechanisms that utilize a visual skin tone signal \cite{fawaz2020skintone} to support the representation of a wide range of skin tones in recommendations, as shown in Figure~\ref{fig:pdp_diversified_small} for fashion recommendations in the \PDP surface. 

 The end-to-end diversification process consists of several components. First, requests that will trigger diversification need to be detected across different categories and locales. Second, the diversification mechanism must ensure that diverse content is retrieved from the large content corpus. Finally, the diversity-aware ranking stage needs to balance the diversity-utility trade-off when ranking content, and to accommodate diversification across several dimensions, such as the skin tone visible in the image as well as the user's various interests. Multi-stage diversification allows the mechanism to operate throughout the pipeline, from retrieval to ranking, to ensure that diverse content passes through all the stages of a recommender system, from billions of items to a small set that is surfaced in the application. 

In this work, we make multiple novel contributions to the area of diverse representation in recommender systems. 
\begin{enumerate}
    \item We present the first visual skin tone diversification production deployment, to the best of our knowledge, to improve representation online in large-scale search and recommender systems.
    \item We developed and productionized a multi-stage diversification system that operates both at retrieval and ranking stages. For ranking, we developed greedy re-rankers and multi-objective optimization using \DPPLong (\DPP), and for retrieval, we implemented a \StrongOR operator for search over token-based indices, as well as \OverfetchAndRerank and \BucketizedANN over embedding-based indices. 
    \item We share learnings from productionizing diversification in a recommender system used by hundreds of millions of users. We also present the challenges, steps, and design choices to mitigate those problems. Our approaches are practical and can easily be translated to other large and complex recommender systems.
    \item We provide empirical results that demonstrate the effectiveness of our various approaches at aligning users' desire for diversity and their utility with the recommendations. 
\end{enumerate}  
As we could increase diversity without negatively impacting utility, and sometimes even increase both, the results suggest that some recommender systems may not be operating at the Pareto frontier between diversity and utility. As diversity at the end of the pipeline is upper-bounded by diversity earlier in the pipeline, the diversity-utility Pareto frontier can be improved by ensuring diversity end-to-end throughout the pipeline, hence the importance of earlier diversification at retrieval.

This paper is organized into five sections -- first, in Section~\ref{sec:problem}, we describe the diversification problem, formulate the diversity metrics, and set up the general mathematical framework. Second, we outline our approach for diversification at the ranking stage of a recommender system by presenting the \RRLong and \DPP-based methods for diversifying a list of Pins with utility scores in Section~\ref{sec:rerank}. Third, in Section~\ref{sec:retrieval}, we present three methods for diversifying at the retrieval layer of a recommender system that alleviate the problem of the lack of diversity at the ranking stage. In Section~\ref{sec:productionization}, we specifically focus on challenges and solutions to enable the practical application of these approaches in a large-scale production recommender system. Finally, in section~\ref{sec:results}, we present the empirical results of our diversity-aware ranking and retrieval methods in terms of diversity and utility metrics in our search and recommender systems. 

In this paper, we detail the methods and results of our research and discuss their implications for the field of recommender systems and the broader tech industry. By addressing the challenge of diversification using practical approaches, it is possible to create more inclusive and equitable products that better cater to diverse communities.

\section{Related work} 
Ranking with novelty and diversity has long been an interest in the field of information retrieval (IR). Diversity is usually defined as a goal for rankings to include different subtopics or cover different possible intents of a query \cite{Carbonell:1998:UMD:290941.291025, Carterette2011, zhu2007improving}. Many methods define evaluation measures for this task, e.g., precision and recall that counts the number of unique subtopics retrieved \cite{zhai2015beyond}, ``information nugget'' based Discounted Cumulative Gain (DCG) that penalizes retrieval of redundant topics \cite{Clarke:2008:NDI:1390334.1390446}, and other generalizations of classical IR metrics to account for diversification \cite{agrawal2009diversifying}. Maximizing these measures is generally NP-hard but the objective functions usually admit a submodular structure \cite{Carterette2011}. Another goal is minimizing the probability of abandonment, or in other words, maximizing the probability of finding the intent in the top-k positions \cite{radlinski2008learning}. \citet{radlinski2009redundancy} distinguish between extrinsic and intrinsic diversity where the former is needed due to the uncertainty in information need (multiple meanings of the query) whereas the latter is needed as a part of the information need. In contrast, diversity for fair representation in recommender systems \cite{geyik2019fairness, zehlike2017fa} is motivated by a different aspect of diversity that is viewed from the perspective of the items being retrieved and ranked, rather than the information need of the user. In this paper, we present the first visual skin tone multi-stage diversification production deployment to improve representation online in large-scale search and recommender systems with hundreds of millions of users and billions of multi-modal Pins.

A wide range of techniques have been leveraged to balance diversity and utility in search and recommender systems \cite{anderson2020algorithmic, wilhelm2018practical, chen2018fast, geyik2019fairness, zehlike2017fa, Carbonell:1998:UMD:290941.291025}. They range from greedy re-ranking heuristics \cite{geyik2019fairness}, potentially using priority queues \cite{zehlike2017fa}, to ranking methods based on pairwise similarity that greedily select results to balance the trade-off between the relevance to the query and the redundancy with respect to previously selected results such as \MMRLong (\MMR) \cite{Carbonell:1998:UMD:290941.291025}, and to multi-objective optimization with probabilistic models such as \DPP \cite{wilhelm2018practical,chen2018fast}. 
 \DPP was found to be more effective at diversifying recommendations than \MMR \cite{chen2018fast} because it accounts for the similarity of all pairs in the union of the selected set and the candidate item, not just the similarity between the candidate item and previously selected item. Solving \DPP is NP-hard. However, thanks to its submodular property, a \DPP solution can be efficiently approximated using a greedy algorithm \cite{gillenwater2012near}, that greedily selects the next item such that the incremental determinant is maximized, and it can be accelerated by updating the Cholesky factor of the \DPP Kernel incrementally \cite{chen2018fast}. 

\section{Multi-Stage Diversification in Search and Recommender Systems}
\label{sec:problem}

\begin{figure}[t]
    \centering
    \includegraphics[width=\linewidth]{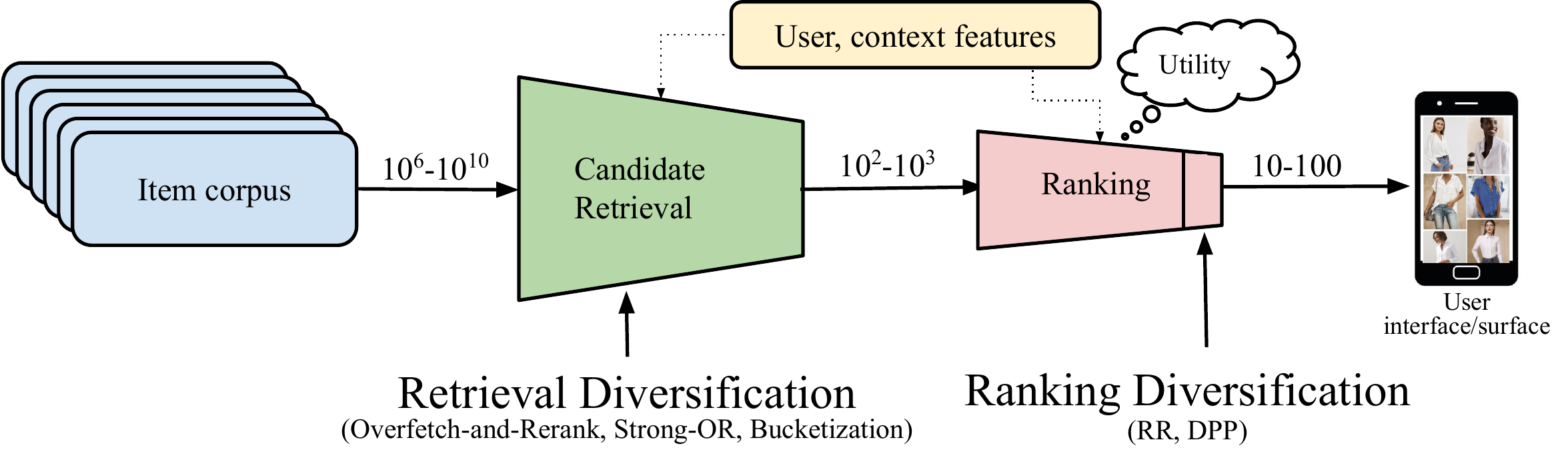}
    \caption{Large-scale recommender systems can broadly be categorized into two stages going from items corpus to recommendations: retrieval and ranking.}
    \Description{This image shows a diagram of the different stages of a large scale recommendation systems. We start with an item corpus of millions or billions of items. The next stage is candidate retrieval, which results in 100s or 1000s of items, and finally ranking, with 10 to 100s of items. For retrieval we can diversify using overfetch and rerank, Strong-OR, or Bucketization. For ranking we can diversify using round-robin or DPP.
}
    \label{fig:two-stage-schema}
\end{figure}

\subsection{Background}
Advanced search and recommender systems, that operate at a large scale with hundreds of millions of active users and billions of items, tend to be very complex and have multiple components. These systems leverage machine learning (ML) models trained to optimize certain objectives given inputs like queries, content features, user features, and past interactions between users and items that happened on the platform. The data in these systems can be multi-modal. For instance, an input query can be in the form of text, such as queries typed by users in a search box; it can be visual when users input an image to search; or it can be a multi-modal item that consists of an image or video, a title, a description, and a link to a website. These systems often comprise two major stages: retrieval and ranking \cite{li2011scene,christakopoulou2018q}, sometimes followed by additional business logic. Items are retrieved and ranked, then the list is surfaced to the user. 

\textbf{Retrieval}: The retrieval stage consists of one or more candidate generators that narrow down the set of candidates from a large corpus of items (in the range of $10^6$ to $10^{10}$) to a much narrower set (in the range of $10^2$ to $10^3$) based on some predicted scores, such as the relevance of the items to the query and the user.  
To achieve high recall and low latency, these systems often leverage search indices, which contain information about each item in a way that enables efficient retrieval. Token-based indices use tokens, e.g., words, as the basic unit of indexing, and are commonly used for text-based search. Embedding-based indices use continuous, dense representations of items to enable retrieval using techniques like approximate nearest neighbor search. These embeddings may be learned through traditional collaborative filtering techniques such as matrix factorization \cite{koren2009matrix} or more advanced methods such as Graph Neural Networks \cite{ying2018graph} or user sequence modeling \cite{pal2020pinnersage}. 
A typical recommender system employs multiple candidate generators, each satisfying different criteria, and their candidates are aggregated before being passed to the next stage.
    
 \textbf{Ranking}: In the ranking stage, the goal is to find an ordering of the items that maximizes an objective or a combination of objectives. 
 The objectives may include utility metrics, diversity objectives, and additional business goals.
 While utility metrics are highly dependent on the application, ranking methods are often simplified to point-wise methods where the first part of ranking is item scoring. The utility scores may be generated by one or multiple ML models trained to optimize certain objectives, such as predicting the probability of an item being relevant to the query, or being clicked, saved, purchased, etc. 
 The second part of ranking is called blending, where multiple objectives are combined to generate a ranked list. A common blending approach for multi-objective optimization is through a weighted combination so that different surfaces can tune or learn weights that best align with their intent. 

\subsection{Diversity in Recommendations}
\label{sec:diversity}
\textbf{Diversity Dimension}: 
Diversification aims to ensure that the ranked list of items surfaced by the system is diverse with respect to a diversity dimension of interest. 
Diversity dimensions may include explicit dimensions such as demographics (e.g., age, gender), geographic or cultural attributes (e.g., country, language), domain-specific dimensions (e.g., skin tone ranges in beauty, cuisine type in food), business-specific dimensions (e.g., merchant sizes), but also other implicit dimensions that may not be expressed directly but can be modeled using latent representations (e.g., embedding, clustering). While this work presents an example of production deployment of skin tone diversification, the proposed techniques are not limited to this single dimension and can support diversification more broadly, including the intersectionality of multiple diversity dimensions. 
We denote the set of groups under a diversity dimension as $\DiversitySet$, and each individual group is denoted by $d_i$ for $i \in \{1,\ldots, |\DiversitySet|\}$.   

\noindent \textbf{Diversity Metric}: 
Given a set of queries $\QuerySet$, we define the top-k diversity of a ranking system $\RankingSystem$ as the fraction of queries where all groups under our diversity dimension, i.e., all $d_i \in \DiversitySet$, are represented in the top $k$ ranked results, denoted by $\RankingSystem_k(q)$. Formally, \metric{k}{\RankingSystem} is defined as  
\begin{align}
    \metric{k}{\RankingSystem} = \frac{1}{|\QuerySet|}\sum_{q \in \mathcal{\QuerySet}} \prod_{d_i \in \DiversitySet} \mathbb{I}[d_i \in \RankingSystem_k(q)] ,
\end{align}
where $\mathbb{I}$ is the indicator function. 
Note that the top-k results $\RankingSystem_k(q)$ are over the items for which the diversity dimension is defined. For instance, in the case of skin tone ranges, a Pin whose image does not include any skin tone would not contribute to visual skin tone diversity. Thus it will not be counted in the top-$k$ and will be skipped in the diversity metric computation.  

In this work, we choose \metric{k}{R} as the primary diversity metric because of its simplicity and intuitiveness as an evaluation metric for the product experience. However, we also compute additional operational metrics to gain a deeper understanding of the effects of diversification on the resulting distribution. For example, the normalized entropy of the distribution with respect to the diversity dimension in top-k items, where the normalization is done against a target distribution. In the case of a uniform target distribution, the metric is called Shannon Equitability \cite{shannon1948mathematical}). Some other metrics proposed in prior work also use divergence measures \cite{geyik2019fairness}. 

\noindent \textbf{Multi-stage diversification:} 
Both retrieval and ranking stages directly impact the diversity of the final content surfaced in the application. The diversity metric at the output of the retrieval stage upper-bounds the diversity at the output of ranking. Hence, the retrieval layer needs to generate a sufficiently diverse set of candidates to ensure that the ranking stage has enough items in each group to generate a final diverse ranking set. However, diversity at the retrieval stage is not a sufficient condition to guarantee that a utility-focused ranker will surface a diverse ordering at the top of the ranking where users are more likely to focus their attention \cite{craswell2008experimental} and to interact with items, especially when such items belong to the long-tail of the distribution \cite{Qin.2021.Long.Tail}. 
Thus, the ranker also needs to be diversity-aware. 

In our ranking diversification experiments on different production surfaces described in Section~\ref{sec:results}, we observed gains in diversity metrics with a neutral to a positive impact on utility metrics, such as user engagement. This suggests that the systems might not have been operating at the Pareto frontier between diversity and utility, as we could increase diversity without negatively impacting utility, and sometimes even increasing both. As we understand that the diversity at the end of the pipeline is limited by the diversity earlier in the pipeline, we can also shift the diversity-utility Pareto frontier further by ensuring diversity end-to-end throughout the pipeline, particularly by introducing diversification earlier at retrieval. Lastly, it should be noted that the retrieval diversity metric is itself limited by the diversity of the content corpus. Diversification at retrieval and ranking cannot correct a lack of representation in the item corpus itself; this can be improved by sourcing more diverse content. 
In Sections~\ref{sec:rerank} and~\ref{sec:retrieval}, we discuss how we introduced diversification at the retrieval and ranking stages in our recommender systems. 

\noindent \textbf{Triggering logic}:  
A real-world system may receive search and recommendation requests that span a wide range of categories, such as fashion, beauty, home decor, food, travel, etc. The diversity dimension of interest depends on the application, for example, skin tone range diversification is applicable to fashion and beauty, but not to home decor. Thus, upon receiving a request, the system needs to determine whether to trigger diversification according to the dimension of interest. 
The triggering logic needs to account for the diversity dimension, the application, the production surface, and the local context, such as country and language, and can be based on heuristics or ML models, such as models that predict the category of a query. 

Diversification and personalization need not be a zero-sum game; they may jointly contribute to an improved user experience. Depending on the application or surface, diversification may be expected in specific dimensions, while personalization may be desirable in others. For instance, for a query related to fashion in \Search or \PDP surfaces, some users may expect to see personalized results in terms of the fashion characteristics of the Pins (e.g., clothing style, fabric, pattern) that relate to those of fashion Pins they have previously interacted with, and yet expect to see a variety of skin tone ranges represented in the human models wearing the fashion Pins in the result images. 
In this case, skin tone diversification allows users to explore fashion images representing a wide range of skin tones rather than a narrower set of results in a single visible skin tone range, creating a sense of inclusion and belonging. 
While some users may expect skin tone diversity in \PDP or \Search surfaces, they may also expect their \Homefeed to be attuned to their interests as reflected by the Pins they chose to interact with, and thus may not expect the same level of exploration and of skin tone diversity in their \Homefeed recommendations as in other surfaces. 
In this work, diversification is triggered for categories and surfaces selected based on user feedback, user research, and data analysis on skin tone related \Search query modifiers that highlight a need for diversity in similar requests. The triggers in this work include beauty and fashion categories in \Search, \PDP, and \NUX (the initial \textit{cold-start} \Homefeed surfaced to a new user). Longer term, one can envision a more advanced system that gives users control over the level of diversification of their results in various contexts through explicit user guidance or learned preferences. 

\section{Diversification at the Ranking stage}
\label{sec:rerank}

We start with a focus on the ranking stage to achieve diversification of results since it is the last stage of a recommender system and it has a direct impact on the metrics we aim to enhance. 
A basic approach to diversify at the ranking stage would be to boost or discount scores for content that is underrepresented according to a diversity dimension. 
While boosters are simple to implement, they tend to add to the technical debt if tuned only at the time they are introduced and if multiple boosters are not optimized together. 
Instead, we leverage a diversity-aware ranking stage that takes as input a list of items with utility scores and their diversity dimensions, and produces a ranking according to a combination of diversity and utility objectives. In this section, we describe two algorithms to achieve diversification through diversity-aware ranking: \RRLong (Section~\ref{sec:grr}) and \DPPLong (Section~\ref{sec:dpp}).

\begin{figure*}[]
    \centering
    \includegraphics[width=\linewidth]{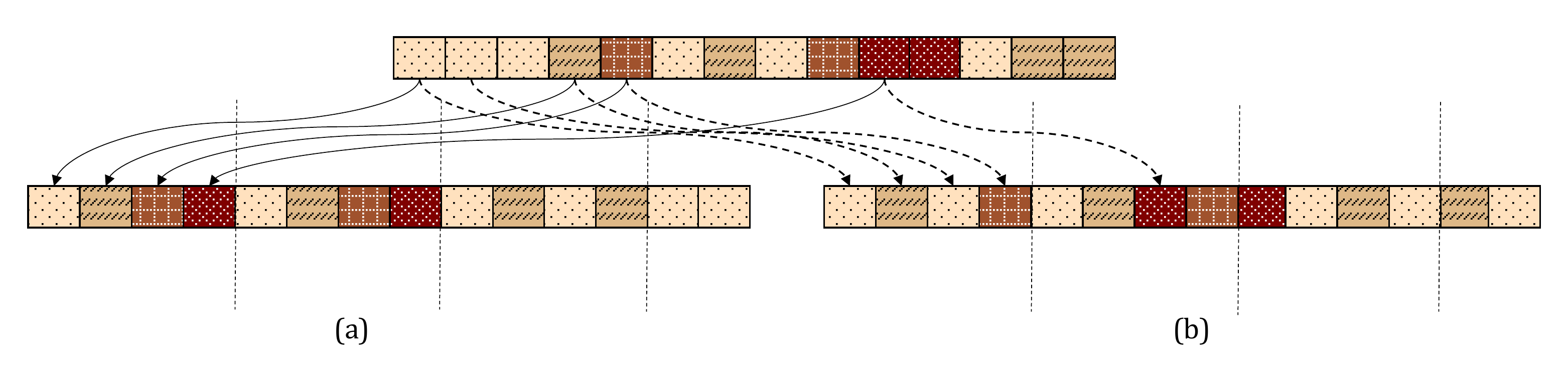}
    \caption{Illustrative example of \RRLong and \DPP applied to a utility-ranked list. Each block is a Pin in the ranked list and the color denotes the skin tone range of the Pin image ($d_i \in \DiversitySet$). (a) Ranked list obtained after applying \RRLong is re-ordered such that the distribution of skin tones is more uniform in the top positions. (b) Ranked list obtained after \DPP (for a specific value of $\theta$) allows for optimizing a list-wise objective to trade off utility and diversity of the initial ranked list. 
    \label{fig:grr_visual}}
    \Description{This image shows an example where round-robin is used to diversify a ranked list of Pins with respect to four groups d1, d2, d3, d4. The initial list has pins in the following order: d1, d1, d1, d2, d3, d1, d2, d1, d3, d4, d4, d1, d2, d2.  With round-robin we get the following order: d1, d2, d3, d4, d1, d2, d3, d4, d1, d2, d1, d2, d1, d1. The distribution of skintones is more uniform in top positions. With DPP we get the following order: d1, d2, d1, d3, d1, d2, d4, d3, d4, d1, d2, d1, d2, d1. 
}
\vspace{-0.15in}
\end{figure*}

\subsection{\RRLong (\RR)}
\label{sec:grr}

The first approach we used is a class of simple greedy rerankers that take in as input a list of items ranked by their utility scores and the item diversity dimension, 
to produce a diversified ranking. Given an ordered list of items $y_1, \ldots, y_n$, we construct $|\DiversitySet|$ number of ordered sub-lists corresponding to diversity dimension and containing items that have a utility score above the threshold. Then, we rebuild a ranked list by greedily selecting the top item of each sub-list one by one. All the items that do not belong to a sub-list, for instance, because they do not have a diversity dimension defined or have utility scores below the threshold, are ranked at the same position as in the original list. 

Figure~\ref{fig:grr_visual} (a) shows an example where \RR is used to diversify a ranked list of Pins with respect to four groups $\{d_1, d_2, d_3, d_4\}$. The list is re-ordered such that the distribution of skin tone ranges is more uniform in the top positions, where users often pay the most attention. The first Pin stays in its original position, and \RR cycles through the skin tone ranges picking the highest-ranked Pin for each sub-list one at a time. A modification to this algorithm could add randomization within a window of size $|\DiversitySet|$ (i.e., 4 in the example) so that, in addition to diversification, it also helps preserve the user experience by ensuring that there is no fixed cyclic order. 

In practical scenarios, the number of items in each sub-list will likely not be even, and \RR may exhaust some sub-lists earlier than others. A few options can be used to handle such cases, such as simply skipping a sub-list when no more items from it are available (e.g., in figure~\ref{fig:grr_visual}(a) when \RR attempts to select the $10^{th}$ Pin but there are no Pins in the $d_3$ sub-list). Another approach could be to merge some of the remaining sub-lists for a more even distribution between lighter and darker skin tones, i.e., merging the sub-lists for $d_1$ and $d_2$, and the ones for $d_3$ and $d_4$, and then continue \RR by alternating elements between the two combined sub-lists.

\RR is a simple, intuitive, and efficient approach to diversification, however, it does not balance diversity and utility and it does not easily generalize to multiple different diversity dimensions or multiple utility score thresholds. To avoid these limitations, in the next section, we describe a multi-objective optimization framework that can balance various utility functions and diversity.

\subsection{\DPPLong (\DPP)}
\label{sec:dpp}

A \DPPLong \cite{macchi1975coincidence, Kulesza.2012.DPP} is a machine learnable probabilistic model used in physics for repulsion modeling and more recently in recommender systems \cite{wilhelm2018practical}. Applications of \DPP range from producing diverse samples of a large database \cite{wilhelm2018practical} to characterizing various observed phenomena like the spatial distribution of fermions in optical beams, where they were originally introduced in \cite{macchi1975coincidence}. DPPs are particularly useful in ML for tasks such as subset selection, where the goal is to select a subset of points from a larger set that is diverse or representative in some sense. In this section, we give a brief overview of DPP (originally introduced in \cite{macchi1975coincidence}) and how it can be applied for diversity-aware reranking.

The basic idea behind a \DPP is to model the probability of selecting a set of items $Y$ from a set of size $N$ as the determinant of a kernel matrix $L_Y$, where $L$ is a kernel function that encodes the utility of the items and the similarity between pairs of items, and $L_Y$ is the kernel matrix of the subset $Y$. The determinant of $L_Y$ can be thought of as a measure of how \textit{spread out} the points in $Y$ are in the feature space defined by the kernel function $L$. The diagonal entry $L_{ii}$ represents the utility of the $i^{th}$ item; in our case the score with which the items were originally ranked. The off-diagonal entry $L_{ij}$, however, represents the similarity between the items, which in our case depends on the diversity dimension, e.g., the skin tone range in the Pin image. The kernel is chosen such that $L$ is a positive semi-definite (PSD) kernel matrix and has a Cholesky decomposition, and hence $L$ can be written as 
\begin{align}
L = U\Phi\Phi^TU^T = USU^T,    
\end{align}
where $U=\mathrm{diag}(e^{\theta u_1}, \ldots, e^{\theta u_N})$ is a diagonal matrix that encodes the \textit{utility} $u_i$ of each item,  $\theta$ is a parameter that governs the trade-off between utility and diversity, and $\Phi = [\Phi_1, \Phi_2, \Phi_3, ..., \Phi_N]$, where $\Phi_i$ is the feature vector for the $i^{th}$ item. For our use case, $\Phi\Phi^T$ is the symmetric similarity matrix, which we henceforth denote by $S$. 
 In terms of set selection in DPP, the probability of selecting a subset $Y$ is proportional to the determinant of $L_Y$.
\begin{align}
  P(Y)\xspace &\propto \xspace \det(L_{Y}) = {{\det}^2(U_Y)} \cdot \det{(S_Y)},\\
  \log \det(L_{Y}) &= 2\theta \sum_{y_i \in Y}{ u_i} + \log \det(S_Y). \label{equation:dpp}
\end{align}

The log determinant is a weighted sum between a utility term and a diversity term balanced by the parameter $\theta$.
Finally, given a value of $\theta$ and kernel matrix $L$, the goal is to find a subset $\overline{Y}$ that maximizes the determinant of $L_Y$:
\begin{align}
    \overline{Y} &=  \argmax_{{Y}\subseteq \{y_1, \ldots, y_N\}}  \det(L_{Y}). \label{eq:set-selection-dpp}
\end{align}
The use of a determinant means that based on the choice of kernel matrix, $\overline{Y}$ would include items with high utility scores while avoiding ones that are similar to others in the subset.
Finding such a subset $\overline{Y}$ of a given size $k$ is an NP-hard problem. However, because of its submodular property, it can be efficiently approximated using a greedy algorithm~\cite{iyer2013submodular}. In the greedy solution, we start with: $Y_0 = \phi$ (empty set), then iteratively add one item at a time to the selected set using the following update rule: 
\begin{align*}
    \overline{Y}_{t+1} &= \overline{Y}_{t} \cup \argmax_{y_i\in \{y_1, \ldots, y_N\}\setminus Y_t} \det\left(L_{\overline{Y}_{t}^w \cup\: y_i}\right) \tag{Greedy Submodular Maximization \cite{chen2018fast}}\\
    &= \overline{Y}_{t} \cup \argmax_{y_i\in \{y_1, \ldots, y_N\}\setminus Y_t} 2\theta { u_i} + \log \det(S_{\overline{Y}_{t}^w \cup\: y_i}) \tag{Using equation~\ref{equation:dpp}}
\end{align*}
where $w$ is a repulsion window size to only consider the last $w$ items (i.e., $\overline{Y}_{t}^w$ in the equation) when computing the $\argmax$. 
The sliding window size $w$ is generally used in applications of \DPP in recommender systems to make the optimization more efficient as the tolerance for similar items may increase as their distance in the ranking increases. 

Figure~\ref{fig:grr_visual}(b) shows a hypothetical example of how \DPP would re-rank as compared to \RR given an appropriate value of parameter $\theta$. Note that setting $\theta$ to a very low value would make \DPP focus primarily on the diversity and hence make the ranking equivalent to \RR. On the contrary, setting $\theta$ to a very high value would make \DPP focus on utility, and hence behave like utility-based ranking. During the implementation, the kernel matrix $L$ can be learned using a deep neural network (e.g., in \cite{wilhelm2018practical}), and $\theta$ can be tuned, e.g., through a grid search with other hyperparameters using offline replay or through A/B experiments. 

In comparison to \RR, \DPP takes into account both the utility scores and similarity and is able to balance their trade-off. For multiple diversity dimensions, \DPP can be operationalized with a joint similarity matrix $S_Y$ to account for the intersectionality between different dimensions. This can be further extended to a function where, for each item, all diversity dimensions (skin tone, item categories, etc.) are provided and the return is a combined value that represents the joint dimensions. A simpler option is to add a diversity term in the weighted sum shown in equation~\ref{equation:dpp} for each dimension. In the case of a large number of diversity dimensions, dimensionality reduction techniques can be used.

\section{Diversification at Retrieval stage}
\label{sec:retrieval}
The ability to diversify in the ranking stage is often limited by the availability of candidates from all groups in the retrieved candidate set. All the techniques proposed in Section~\ref{sec:rerank} are limited to the set of candidates retrieved by the different candidate sources in the first stage, and hence, for specific queries, it may not be possible to diversify the ranking at all in the ranking stage. To tackle this limitation, we propose a set of techniques that increase the diversity of candidates at the retrieval layer to enhance the ability of the re-rankers to diversify at a later stage. This section presents three techniques: \OverfetchAndRerank, \StrongOR logic, and \BucketizedANN. While \OverfetchAndRerank is generally applicable for any retrieval stack, it has certain limitations that can make it impractical for many real-world diversification scenarios. Hence, we propose \StrongOR and \BucketizedANN to tackle these limitations for two specific use cases of retrieval.

\subsection{\OverfetchAndRerank at Retrieval}
\label{sec:overfetch}

One of the simplest ways to increase the diversity of the candidate set at retrieval is to fetch a candidate set of a larger size (\textsc{Overfetch}). In this case, we may define the desired diversity criterion as the property that the candidate set contains a minimum threshold number of candidates from each group. For example, if we want to retrieve a candidate set of size~$K$ through K-nearest neighbor search in an embedding space, we could expand the neighborhood size to $K'$ nearest neighbors ($K'>K$) such that the resultant candidate set has at least $k$ candidates from each group. 
As the next step to only pass $K$ candidates to the ranking stage, we can perform a \RRLong selection of a subset of size $K$ from this over-fetched set of size $K'$, for example, selecting one candidate at a time from each skin tone range until $K$ candidates are selected (\textsc{Rerank}).
However, as expected, the expanded size of the neighborhood $K'$ to be explored is limited due to the increase in latency that the retrieval stage can afford. Hence, we choose a hyperparameter $\kmax$ such that $K'$ never exceeds $\kmax$. The overfetching will stop when either the minimum threshold in each group is met or when $K' = \kmax$. In Appendix~\ref{sec:choice-of-kprime}, we discuss the choice of this parameter.

\subsection{\BucketizedANN}
\label{sec:bucketized}
One of the most commonly used retrieval methods in an embedding-based search index is the approximate nearest neighbor (ANN) search. For embedding-based retrieval, the users, items, and queries are all embedded into the same space, and for applications like search and recommender systems, the system wants to retrieve the items that are closest to the query or user embedding in terms of a chosen distance metric (e.g., cosine distance). Since computing pairwise distances for all query-item pairs is prohibitive in a practical recommender system, this nearest neighbor search is often performed using approximation algorithms that rely on efficient data structures, e.g., k-Dimensional Tree \cite{bentley1975multidimensional}, Locality-sensitive Hashing (LSH) \cite{indyk1997locality}, and Hierarchical Navigable Small Worlds (HNSW) \cite{malkov2018efficient}. In this work, we will refer to these methods as ANN search methods. Most of these approximation algorithms partition the embedding space into multiple regions and perform a search in it. In larger-scale recommender systems where this search is over billions of items, these ANN methods are implemented as a distributed system, like \citet{yianilos1993data} for example. 

\begin{figure}
    \centering
    \includegraphics[width=\linewidth]{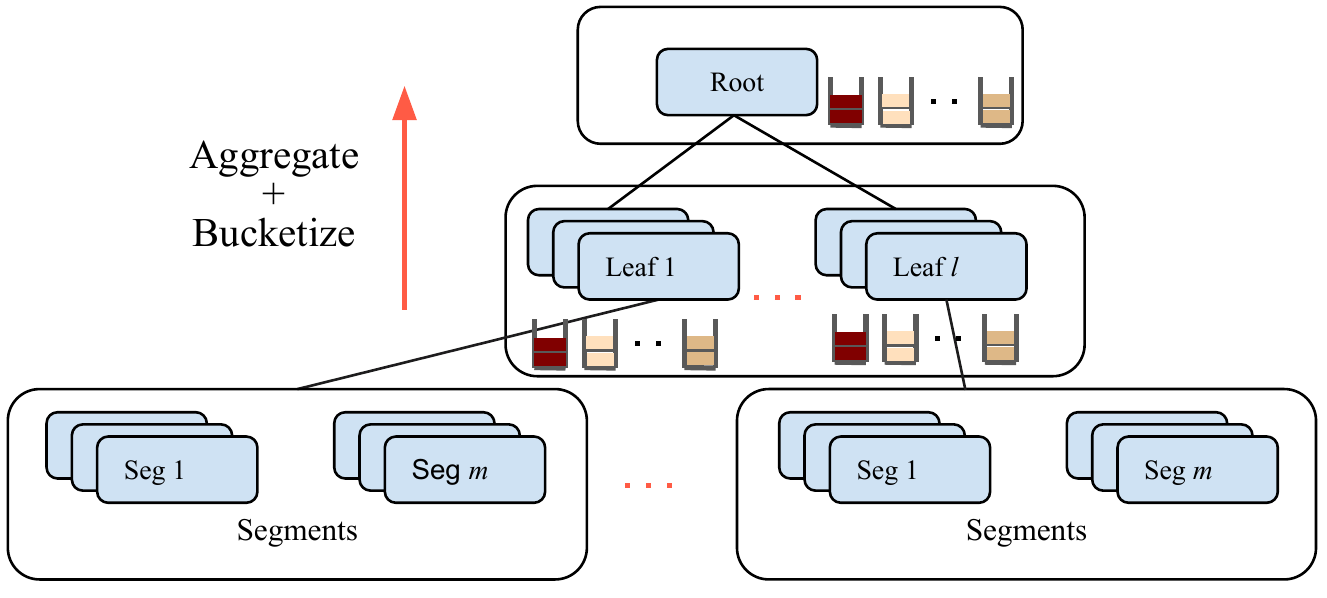}
    \caption{A diagram of distributed ANN retrieval aggregating candidates from segments to leaves to the root based on the distance metric while assigning top Pins with each skin tone to their corresponding buckets. }
    \Description{A diagram of distributed ANN retrieval aggregating candidates from segments at the bottom to leaves in the middle and the root at the top based on the distance metric while assigning top Pins with each skin tone to their corresponding buckets.
}
    \label{fig:ann-retrieval}
\end{figure}

For this work, we will follow the general architecture of an ANN search system that contains a root node that sends a request to a few leaf nodes that further request several segments to perform a nearest neighbor search in different subregions of the embedding space (as shown in Figure~\ref{fig:ann-retrieval}). Let's say there are $L$ number of leaves and $M$ number of segments per leaf; to find $K$ nearest neighbors for a given query embedding, each segment returns $K$ potential nearest neighbor candidates to the corresponding leaf, which then aggregates these $M \times K$ number of candidates to only retain the top $K$ candidates, before passing it along to the root. The root is then responsible for choosing the top $K$ candidates from the $K \times L$ candidates whose exact distances are computed during the process. Note that the size of the graph, in this case, is $K \times L \times M$.

For the \BucketizedANN approach, we modify the aggregation step (at the leaf and the root level) to also aggregate top $K_{d_i}$ candidates from each group  $d_i\in\DiversitySet$ into buckets corresponding to each of the groups under the diversity dimension (in addition to aggregating the top-$K$ candidates in the overall pool). In other words, each leaf now aggregates a set of $K$ candidates, and $|\DiversitySet|$ buckets with (at most) $K_{d_i}$ candidates each. This helps preserve the top candidates belonging to each group (whose distances are already computed) from being dropped during the aggregation steps, without incurring the high cost of expanding the entire aggregation graph in the \OverfetchAndRerank approach.

\begin{figure}
    \centering
    \includegraphics[width=\linewidth]{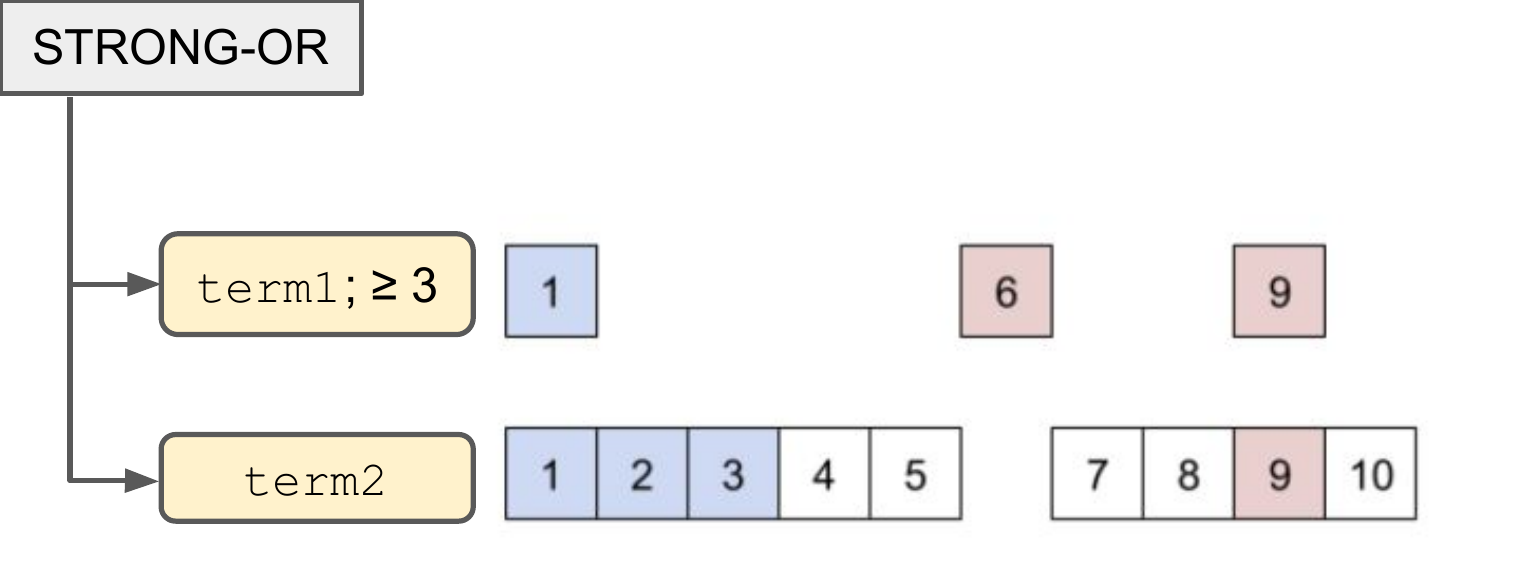}
    \caption{An example of how the \StrongOR operator ensures diversity during retrieval for two query terms, one with a minimum threshold condition.}
    \label{fig:strong_or}
    \Description{The image shows an example of a query containing the Strong Or syntax. The query contains two terms under the Strong Or logical operator: term 1 and term 2. The query also specifies a condition for candidates matching term 1 that there must be greater than or equal to 3 candidates that satisfy term 1. In terms of candidates available, there are only three candidates matching term 1, namely 1, 6 and 5, and 9 candidates matching term 2, i.e., all candidates except 6. The candidates are ordered in the ascending order already, i.e., candidate 1 is more relevant to the query than candidate 2 and so on. To satisfy the condition, the Strong Or operator now selects all the three candidates under term 1, and since candidates 1 and 9 also satisfy term 2, the operator only has to pick candidates 1, 2 and 3 to complete the set of 5 candidates.}
\end{figure}

\subsection{\StrongOR Retrieval}
\label{sec:strongOR}
In \Search, one of the critical components of the retrieval stage is query understanding, where the text query is converted to a structured query (s-query). It allows the retrieval system to specify relationships between different query terms using logical operators (such as AND, OR, XOR) to connect tokens in order to narrow down or broaden the set of results, e.g., a text query may be parsed to \texttt{dress AND (red OR black)}. Since we would like to broaden the set of results to contain candidates with underrepresented groups under the diversity dimension, we use a specialized logical operator for search called \StrongOR \cite{curtiss2013unicorn}. 

On a search index, \StrongOR operates similarly to the OR operator except that it prioritizes a candidate set that satisfies multiple criteria simultaneously. More concretely, like an OR operator, we can specify that the retrieved candidates must belong to either $d_1$ or $d_2$, but in addition, we can also specify what (minimum) percentage of candidates match each of the respective criteria. In an s-query, \StrongOR expresses the disjunction semantics across all children of the parsed query (e.g., Figure~\ref{fig:strong_or}), and each child node in the s-query can optionally be required to match $\geq k_d$ of all candidates. If there are insufficient candidates to fulfill the criteria specified, it will match as many as possible. 

Given an early stop parameter $K$ limiting the maximum number of candidates to be scanned rather than scanning the entire corpora, a criterion denoted by $\Gamma$ (e.g., in Figure~\ref{fig:strong_or}, $\Gamma$ is \texttt{term1} $\geq p\%$, $k_d=5$ and $K=10$), \StrongOR fetches $K$ candidates that match the query. As we scan the list from left to right, \StrongOR acts as a regular OR at first, i.e., if it naturally satisfies $\Gamma$, the set is returned. Otherwise, during the scanning, \StrongOR promotes $\Gamma$ to be a required criterion; for example, in Figure~\ref{fig:strong_or}, after position 3 in the list, $\Gamma$ becomes a necessary condition, and hence candidates 6 and 9 (in red) are retrieved instead of 4 and 5. Since \StrongOR happens in the retrieval stage during the query understanding phase, we can also add the candidates, that satisfy $\Gamma$ and would not have been retrieved otherwise, into dedicated buckets (like in \BucketizedANN) to ensure that they are not dropped in the latter stages of retrieval.  

\section{Productionization Considerations for a large scale recommender system}
\label{sec:productionization}
We implemented and deployed diversification approaches in a large-scale recommender system with over 460 million monthly users \cite{pinterestEarnings}, who use Pinterest to find visual inspiration for their interests and simultaneously find products that fit their needs. We chose three different surfaces on the platform based on user feedback to diversify specific experiences -- namely \Search, \NUX, and \PDP. On \Search, users enter a text query to find Pins that match their intent. The \PDP surface recommends a list of Pins similar to the Pin selected by the user (query Pin). All Pins recommended on this surface are products\footnote{Pinterest contains over 1 billion product Pins available for recommendations.} available for purchase by users. Lastly, \NUX is the initial \Homefeed that a new user sees after signing up. These surfaces were consciously chosen keeping in mind user research and data analysis of user needs as mentioned in Section~\ref{sec:diversity}. 
In the rest of this section, we present multiple practical considerations to deploy our diversification approaches in a real-world production system. 

\subsection{Indexing}
Deploying diversification algorithms at retrieval requires indexing the diversity dimension of Pins, e.g., the Pin skin tone range, in both embedding-based and token-based indices. 
For each Pin, the skin tone range (if applicable) is computed offline using a computer vision model. An offline batch workflow periodically reads the skin tone predictions generated for each Pin from a store and adds it to the indexing pipelines of each surface for fast retrieval. The indexed diversity dimension can be passed along with the candidate Pins to the ranking stage for ranking diversification. Alternatively, the ranking stage can read the Pin diversity dimension from stores or a caching service. 
Once the Pin diversity dimensions are available in the serving infrastructure, we implement various diversification algorithms at retrieval and ranking stages and run online experiments to assess their performance in different geographic markets. 

\subsection{Latency and Scaling Considerations}
One of the main advantages of using \RR is its implementation simplicity as a post-ranking step, even for a complex recommender system, and its minimal impact on latency due to the linear time complexity. However, because of the use of sub-lists for each group, it is hard to scale when new diversity dimensions need to be incorporated. For example, if we desire to diversify for skin tone as well as the category of the Pins (e.g., home decor, fashion, beauty), the number of possible combinations may become impractical for \RR. A possible solution to this problem is to use priority queues to iterate over these multiple dimensions. Another downside of \RR is that it does not clearly balance the trade-off between diversity and utility scores. We also need to define how the diversification algorithm handles Pins that do not belong to a specific group in the diversity dimension, for example, images that do not contain skin do not have a skin tone. These cases can be handled by either leaving them at their original position and only diversifying the Pins where a skin tone is detected, or by randomly sampling and assigning a diversity dimension to those Pins and having them be part of \RR. 
Lastly, it may be beneficial to add a threshold of how deep we consider Pins for \RR, either by position or the utility score. Limiting the set of Pins we perform \RR over helps ensure a lower impact on the overall latency. In addition, it also takes into consideration that Pins that were originally ranked lower may be less relevant to users. That being said, it is crucial to assess the fairness of the ranking models before thresholds are used as Pins from certain diversity dimensions may be disproportionately ranked higher or lower by some ML models.

For \DPP, the similarity matrix is computed at serving time for the list of Pins in the ranking. While the greedy iteration computes the diversity term at each step, the utility scores can be cached during the DPP iterations.  Given that the time complexity of DPP is $O(wBN)$ \cite{chen2018fast}, we apply a few techniques to reduce the impact on latency, that can be optimized and evaluated through offline replay, shadow testing, or A/B experiments for each surface:
\begin{itemize}[leftmargin=2em]
    \item Tuning the batch size and the window size: Given a list of $N$ utility-ranked Pins, instead of \DPP re-ranking the entire list of size $N$, we can diversify the ranking of Pins only up to a certain position \DPPBatchSize (the batch size) or above a certain score threshold $\tau_B$, over sliding windows of size $w<N$. For example, one may only consider diversifying the median (or a percentile above the median) scrolling length in the surface. 
    \item Tuning the depth size: While diversifying a batch of size \DPPBatchSize, we can seek diverse Pins all the way to the depth size \DPPDepth where $\DPPBatchSize \leq \DPPDepth \leq N$ to generate \DPPBatchSize diversified results. Both the batch size and the depth size reduce the computation required to rerank, but the increase in diversity in top-k could be limited by the availability of diverse content in the explored depth. 
    \item Batch parallelization: Building on the concept of batch size, for some surfaces where users scroll deeper, we can also diversify multiple batches of size \DPPBatchSize. This way the set of Pins is diversified throughout the ranking while limiting the initial loading time for the first set of Pins. 
\end{itemize}

\subsection{Qualitative evaluations}
\label{sec:qualitative}
To evaluate the diversification of results using skin tone, we collected qualitative feedback from a diverse set of internal participants for every iteration. We presented them with a side-by-side comparison of results before and after diversification for different conditions in our A/B experiments and asked them to rate the results based on diversity and relevance. We also collected relevance evaluations through professional data labeling, where raters evaluated the relevance of results for a sample of queries. Additionally, to account for the local context in international markets, we collaborated closely with the internationalization team for a qualitative assessment of diversification and its results in various markets. These inputs were extremely important in making the final decision to launch our approaches in international markets.

\section{Results in Production}\label{sec:results}

To study the impact of diversification on business metrics, including user engagement, and the impact on the diversity metric, we ran several A/B experiments on three surfaces on the Pinterest app: \Search (for fashion and beauty-related queries), \PDP (for fashion Pin recommendations), and \NUX (fashion and beauty category Pins). 

There are several nuances that must be taken into consideration when measuring the success and implications of these approaches in search and recommender systems. First, appropriate metrics and guardrails must be set in place before performing diversification. Second, while some of the learnings are transferable between surfaces, each surface presents unique challenges and may differ drastically from past use cases. The differences between surfaces encompass but are not limited to active users, Pin corpora, business metrics, and surface goals. Because of these factors, comparing the change in the diversity metric between surfaces is a moot point as even the data used in each surface is often different and sometimes disjoint. 
Nonetheless, we often observed positive gains in diversity metrics coupled with neutral or positive impact in guardrail business metrics for all the techniques described in this paper. 
It is also worth noting that not all surfaces require the same types of interventions so techniques described in earlier sections were applied to surfaces when appropriate.  

The metrics reported in this paper are the result of several A/B experiments we ran in production for at least 3 weeks in the US as well as international markets. The number of users varies per surface, \Search and \PDP both had a few million users per experiment group, while \NUX had hundreds of thousands of users per group. In the rest of this section, we give a brief overview of the impact of these techniques on user engagement metrics and the diversity metric ($\metric{k}{R}$) (we provide more details on the choice of $k$ in Appendix~\ref{sec:choice-of-k}). We report the impact to these metrics as the percentage difference relative to control. Any impact on engagement metrics reported below was statistically significant with p-value < 0.05. 

\subsection{\Search}
\begin{figure*}
\centering
\begin{subfigure}{.3\textwidth}
\centering
\includegraphics[width=5cm,height=5cm,keepaspectratio]{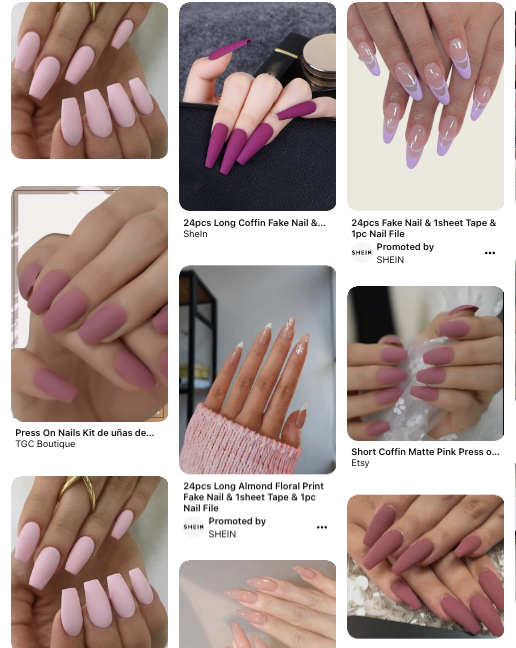}
\caption{Control}\label{fig:control-nails}
\end{subfigure}%
~
\begin{subfigure}{.3\textwidth}
\centering
\includegraphics[width=5cm,height=5cm,keepaspectratio]{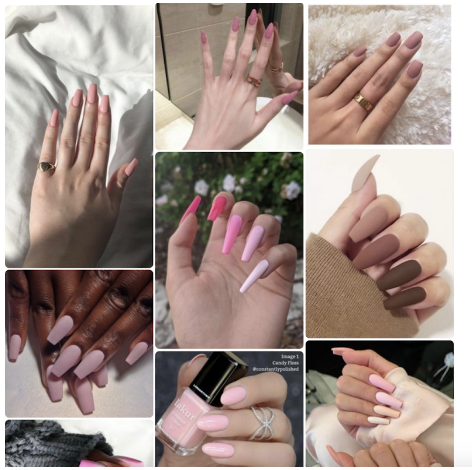}
\caption{\RR}\label{fig:rr-nails}
\end{subfigure}%
~
\begin{subfigure}{.3\textwidth}
\centering
\includegraphics[width=5cm,height=5cm,keepaspectratio]{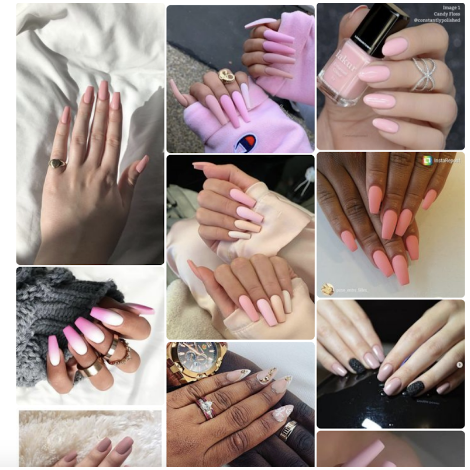}
\caption{\DPP}\label{fig:dpp-nails}
\end{subfigure}%
\vspace{-0.15in}
\caption{For the query ``pink nails matte'' on \Search, (a) shows search results without any diversity, (b) shows diversified search results using \RR with a score threshold, and (c) shows the diversified ranking for the same query using \DPP.}
\Description{For the query “pink nails matte” on Search, the first image shows search results without any diversity, which has a narrow range of skin tones. The second image shows diversified search results using RR with a score threshold and the third image shows the diversified ranking for the same query using DPP. The last two images have a wider range of skin tones represented across search results.}
\label{fig:search_diversified_1}
    \vspace{-0.15in}
\end{figure*}

In order to diversify search results with respect to skin tones, we first adopted \RR with a score threshold for queries in the beauty and fashion categories in an A/B experiment that ran for five weeks in October 2020. This approach led to a 250\% increase in the diversity metric and had a positive impact on engagement. We further iterated on the ranking stage by replacing \RR with \DPP in an experiment that ran for three weeks in April 2021, which resulted in a minor impact on the diversity metric while improving engagement and user growth metrics. The \DPP approach was also launched in some international markets where we saw a similar trend in metrics, with gains in the diversity metric ranging from 200\% to 400\% for different countries, with respect to the non-diversified baseline. We also observed an improvement in daily active users (number of users who made at least one request to Pinterest from any device), weekly active users (derived from daily active users by looking at action type for at least 7 days), and overall time spent on the platform. 

To diversify in the retrieval stage, we deployed the \StrongOR logic to improve the diversity of the Pins retrieved. The experiment was run for a period of four weeks in August 2022. Adding this retrieval logic along with existing \DPP at ranking resulted in an additional 14\% increase in the diversity metric as well as a positive impact on search engagement. Figure~\ref{fig:search_diversified_1} shows a visual comparison of search results before and after diversification.

\subsection{\PDP}
In \PDP, after assessing the various cases and Pin categories where it would be appropriate to start deploying diversification, we introduced diversification in the ranking and retrieval stages for fashion and wedding-related Pins. For ranking stage diversification, we initially ran an A/B experiment with \RR to diversify results post-ranking for the treatment group. This experiment was run for a period of ten weeks between September and November 2020. We observed a 270\% improvement in the diversity metric $\metric{k=10}{\RankingSystem}$ and a neutral impact on relevance and engagement for the treatment group compared to the control group. With the increase in diversity in the top-ranked recommendations, we observed that the diversity also increased in terms of the skin tone distribution of the Pins that users engaged with. This experiment led to a successful launch of \RR to all users viewing the fashion and wedding Pins. 

To better balance ranking scores and diversity, we conducted another A/B experiment comparing \RR (as control) to \DPP (as treatment) for a period of four weeks during March and April 2022. Since \DPP tries to balance utility scores of the Pins with diversity as compared to \RR that reorders Pins only on the basis of the diversity dimension, as expected, the treatment group saw a small decrease in the diversity metric $\metric{k=10}{\RankingSystem}$, while some shopping metrics like the proportion of users with purchases increased by 1.3\%, and engagement metrics - such as clicks, long-clicks and saves - increased by more than 5\%. At the end of the experiment, \DPP was deployed to all users as part of the default \PDP experience.

To introduce diversification at the retrieval stage, we implemented the \BucketizedANN techniques to enhance the diversity of the retrieved set. In an experiment (that ran for 11 weeks between December 2022 and March 2023), \BucketizedANN led to an increase in the diversity metric (of the entire candidate set) by 8\% at the retrieval stage for the nearest neighbor search in a Pin embedding space, while the relative increase in the diversity metric was about 1\%. The relatively small increase due to retrieval diversification means that more work is needed to tune the hyperparameters at the ranking stage with respect to the diverse retrieval set. 

Prior to each launch, as outlined in Section~\ref{sec:qualitative}, we conducted qualitative evaluations to compare the relevance of top-ranked recommendations and observed no significant changes in relevance or recommendation quality.  Figure~\ref{fig:pdp_diversified_small} shows a visual comparison of \PDP results before and after \DPP-based ranking combined with \BucketizedANN-based diversification. 

\subsection{\NUX}

We introduced diversification as part of the new user experience so that everyone feels represented from their first interaction with the platform. We initially developed a two-dimensional \RR variation, which prioritized Pin category diversity using a category \RR and achieved best-effort skin tone diversity using a priority queue. Leveraging a frequency-based skin tone priority queue, it greedily selected the next Pin at each step in the re-ranking, so that skin tone ranges with lower frequency across topics were given higher priority. In an A/B experiment run over six weeks in April to June of 2021, we deployed this approach for a subset of skin tone-related categories within the beauty and fashion categories. The skin tone diversity metric increased 109\% with a neutral impact on Pin category diversity, engagement, and growth metrics. 
Iterations on our ranking system led us to replace the two-dimensional \RR with a single skin tone based \RR in the ranking stage, that operated over all Pins with a skin tone range across categories. 

This experiment that ran for four weeks in September 2021 led to a 650\% improvement in the diversity metric as compared to the non-diversified experience with a neutral impact on engagement. 
Finally, we introduced diversification at the retrieval stage using \OverfetchAndRerank in an experiment that ran for four weeks in March 2022, which increased the skin tone diversity metric by 63\%, with a neutral impact on latency. 
In the latest deployment, \DPP diversification was introduced in the ranking stage (the experiment was run for four weeks in November 2022), leading to a 462\% increase in the diversity metric due to retrieval diversity alone. 
Ultimately, the combination of \DPP and \OverfetchAndRerank achieved the best balance in terms of diversification and utility for \NUX.

\section{Ethical considerations}
\label{sec:ethical}

Skin tone diversification aims at improving representation by surfacing all skin tone ranges in the top results when possible. While the visible skin tone ranges in Pin images are leveraged to surface all skin tone ranges in the top results at serving time, they are not used as inputs to train ML ranking models. It is important to note that skin tone ranges are Pin features, not user features. We respect the user's privacy and do not attempt to predict the user's personal information, such as their ethnicity.  


\section{Conclusions and Future work}
We addressed the challenge of diversification to improve representation in search and recommender systems using scalable diversification approaches at ranking and retrieval. We deployed multi-stage diversification in a large-scale production system with hundreds of millions of users, and through extensive empirical evidence showed that it is possible to create an inclusive product experience that positively impacts utility metrics such as engagement. We shared the learnings from deploying the first visual skin tone diversification in a visual discovery recommender system to the best of our knowledge. Our techniques are scalable for multiple simultaneous diversity dimensions and can support intersectionality. 

Future work includes developing more advanced and scalable triggering mechanisms, for instance through a model that learns which requests to diversify for the diversity dimensions of interest based on the context, including the query, surface, category, and language.  
 In ranking, the multi-objective optimization weights that balance different objectives including diversity could be adapted over time in an automated manner. Future work in retrieval diversification can take inspiration from recent research in debiasing word embeddings \cite{bolukbasi2016man} and fair representation learning \cite{zemel2013learning} to ensure that the underlying representations in embedding-based retrieval are fair for relevant diversity dimensions. We can analyze how diversified search results and recommendations can help mitigate serving bias in systems that generate their own training data, by creating a positive feedback loop for model retraining thanks to richer interaction data from a diverse set of Pins. Finally, we can evaluate the potential impact on the diversity of Pins in the corpora in the long term.

\begin{acks}
    This endeavor would not have been possible without several rounds of discussion and iterations with our colleagues --
    Vinod Bakthavachalam, Somnath Banerjee, Kevin Bannerman-Hutchful, Josh Beal, Larkin Brown, Hayder Casey, Yaron Greif, Will Hamlin, Edmarc Hedrick, Felicia Heng, Dmitry Kislyuk, Anna Kiyantseva, Tim Koh, Helene Labriet-Gross, Van Lam, Weiran Li, Daniel Liu, Dan Lurie, Jason Madeano, Rohan Mahadev, Nidhi Mastey, Candice Morgan, AJ Oxendine, Monica Pangilinan, Susanna Park, Rajat Raina, Chuck Rosenberg, Marta Scotto, Altay Sendil, Julia Starostenko, Kurchi Subhra Hazra, Eric Sung, Annie Ta, Abhishek Tayal, Yuting Wang, Dylan Wang, Jiajing Xu, David Xue, Saadia Kaffo Yaya, Duo Zhang, Liang Zhang, and Ruimin Zhu. 
    We would like to thank them for their support and contributions along the way.
\end{acks}

\bibliographystyle{ACM-Reference-Format}
\bibliography{inclusiveAIrefs}

 \appendix
 \section{Choice of hyperparameters} 
 \label{sec:hyperparams}
We tune the hyperparameters for \RR, \DPP, \StrongOR, and \OverfetchAndRerank using both online experiments and offline evaluations. 

 \subsection{\RRLong hyperparameters}
In \RR, score thresholds allow mitigating potential impact on utility metrics. We evaluated different values of the score threshold to determine which Pins should be included in the \RR logic. The Pins that did not meet the threshold were appended towards the end in the same order they were ranked in the utility-based ranking list. 

\subsection{\DPPLong hyperparameters}
For \DPP, we iterated on the different parameters like window size $w$, utility-diversity trade-off parameter $\theta$, batch size \DPPBatchSize, and the similarity function (Sections~\ref{sec:productionization} \& \ref{sec:dpp}). \DPPWindowSize is usually based on how much user fatigue there is when similar Pins show close to each other. If $|\DiversitySet|$ is relatively small it is possible to set this value to $|\DiversitySet|$ itself or multiples of it, but for large values of $|\DiversitySet|$ further testing is necessary to properly set $w$. The batch size is the position up to which we diversify the ranking (instead of the entire candidate set) to reduce the computation during DPP. We experimented with batch sizes of 200, 400, and 800. Next, the value of $\theta$ determines how much weight we want to give to the utility score as compared to diversity, if $\theta$ is increased by a lot, the greedy optimization of DPP becomes unstable. For kernel transform, we tried two types of kernel: Identity and Radial basis function (RBF). The latter performed better on our data in terms of diversity. 
Lastly, for the similarity function, we experimented with mathematical variations of exponential, linear, and cosine similarities when comparing the skin tone ranges of two Pins. Through our experiments, we find that in several surfaces, the linear similarity performed closer to \RR in terms of diversity, however the choice of this parameter may vary for each surface, given the Pin corpus varies between them.  

\subsection{\OverfetchAndRerank hyperparameters}
\label{sec:choice-of-kprime}
We experimented with various values for the maximum overfetching parameter $\kmax$ for \OverfetchAndRerank in \NUX. The results revealed that setting the overfetching upper-bound $\kmax$ to twice the candidate size offered a good balance between latency and diversity for the surface.

\subsection{Choice of \boldmath$k$\unboldmath~in~\boldmath\metric{k}{R}\unboldmath}
\label{sec:choice-of-k}

\begin{table}
\begin{tabular}{@{}ccc@{}}
\toprule
k  & $\Delta(\metric{k}{R_{\text{prod}}})$ & $\Delta$(num queries) \\ \midrule
6  & -44.54\%            & +36.76\%                   \\
8  & -17.77\%            & +15.57\%                   \\
10 & \multicolumn{2}{c}{$\mathrel{\vcenter{\hbox{\rule{1cm}{0.5pt}}}}Baseline \mathrel{\vcenter{\hbox{\rule{1cm}{0.5pt}}}}$}\\
15 & +27.78\%            & -25.65\%                 
\end{tabular}
\caption{Comparing the relative change in the diversity metric and number of queries for different choices of $k$ based on the impression log. For this work, we choose $k=10$ as the depth of ranking in our metrics.\label{table:k-v-metric}} 
\end{table}

We consider two factors when choosing a value of $k$ for the diversity metric \metric{k}{R}: maximizing the coverage in terms of queries, and observing perceptible changes in the diversity of the rankings.
In \Search and \PDP surfaces, we choose $k=10$ in the \metric{k}{R} and $k=6$ for \NUX to balance the two factors. While choosing a lower value of $k$ makes it harder to satisfy a diversity constraint, choosing a higher value of $k$ reduces the amount of data we can use to compute the diversity metric because not all users view impressions up to a large $k$. To make this decision, we collected Pin impressions logs from the recommendation surfaces to study the relationship between $k$ and our metric $\metric{k}{R}$ for our production ranking system $R_{\text{prod}}$. In Table~\ref{table:k-v-metric}, we show some of these values for the \PDP surface: the relative change in the diversity metric (fraction of queries with all the skin tone ranges represented in the top-k) and the relative change in the number of queries with at least $k$ Pin impressions that contain a skin tone range.
As expected, choosing a higher value of $k$, increases the diversity metric but reduces the number of queries (or requests) used to compute the metric itself which may lead to differences in the metric value. Hence, we decide to choose $k=10$ for this surface as there is sufficient scope for improvement in the diversity metric possible within the first ten Pins while we also have enough number of queries to compute the metric. 

\end{document}